\pdfoutput=1

\documentclass[11pt]{article}
\usepackage{multirow}
\usepackage{amssymb}
\usepackage[]{acl}
\usepackage{colortbl}
\usepackage{times}
\usepackage{latexsym}
\usepackage{url}

\usepackage{enumitem}
\usepackage[T1]{fontenc}

\usepackage[utf8]{inputenc}

\usepackage{microtype}
\usepackage{inconsolata}

\usepackage{graphicx}
\usepackage{tikz}
\usepackage{subcaption}
\newcommand{\circled}[1]{\tikz[baseline=(char.base)]{
            \node[shape=circle,draw,inner sep=1pt] (char) {#1};}}

\title{DocSpiral: A Platform for Integrated Assistive Document Annotation through Human-in-the-Spiral}

\author{
  \textbf{Qiang Sun\textsuperscript{1}\textsuperscript{*}},
  \textbf{Sirui Li\textsuperscript{2}},
  \textbf{Tingting Bi\textsuperscript{3}},
  \textbf{Du Huynh\textsuperscript{1}},
  \textbf{Mark Reynolds\textsuperscript{1}},
  \textbf{Yuanyi Luo\textsuperscript{4}},
  \textbf{Wei Liu\textsuperscript{1}\textsuperscript{*}}
\\
  \textsuperscript{1}The University of Western Australia, Perth, WA, Australia,
  \textsuperscript{2}Murdoch University, Perth, WA, Australia, \\
  \textsuperscript{3}The University of Melbourne, Melborune, VIC, Australia, \\
  \textsuperscript{4}Sinograin Chengdu Storage Research Institute Co., Ltd., Chengdu 610091, China
\\
  \small{
    \textsuperscript{*}\textbf{Correspondence:} \href{mailto:pascal.sun@research.uwa.edu.au}{pascal.sun@research.uwa.edu.au}, \href{wei.liu@uwa.edu.au}{wei.liu@uwa.edu.au}
  }
}

\begin{document}
\maketitle
\begin{abstract}
Acquiring structured data from domain-specific, image-based documents—such as scanned reports—is crucial for many downstream tasks but remains challenging due to document variability. 
Many of these documents exist as images rather than as machine-readable text, which requires human annotation to train automated extraction systems.
We present \textbf{DocSpiral}, the first Human-in-the-Spiral assistive document annotation platform, designed to address the challenge of extracting structured information from domain-specific, image-based document collections.
Our spiral design establishes an iterative cycle in which human annotations train models that progressively require less manual intervention. 
\textbf{DocSpiral} integrates document format normalization, comprehensive annotation interfaces, evaluation metrics dashboard, and API endpoints for the development of AI / ML models into a unified workflow.
Experiments demonstrate that our framework reduces annotation time by at least 41\% while showing consistent performance gains across three iterations during model training.
By making this annotation platform freely accessible, we aim to lower barriers to AI/ML models development in document processing, facilitating the adoption of large language models in image-based, document-intensive fields such as geoscience and healthcare. The system is freely available at: \url{https://app.ai4wa.com}. The demonstration video is available: \url{https://app.ai4wa.com/docs/docspiral/demo}.
\end{abstract}

\section{Introduction}
Unstructured data are information that does not adhere to a predefined data model or format, such as free text, images, audio, video, and social media content, etc.~\cite{techtarget_unstructured_data}.
Unstructured data are widely believed to form 80\%-90\% of the world's global data assets~\cite{researchworld_unstructured_data}. 
Due to the sheer complexity of dealing with such data, unstructured data are often referred to as ``dark data'' and are significantly underutilized.
To unlock the wealth of valuable knowledge hidden in unstructured data, various techniques such as knowledge graph constructions and Retrieval Augmented Generation~(RAG~\cite{stewart2020seq2kg}) can be employed, provided that the documents are first processed to extract relevant textual data.

\begin{figure}[hbt]
    \centering
    \includegraphics[width=\linewidth]{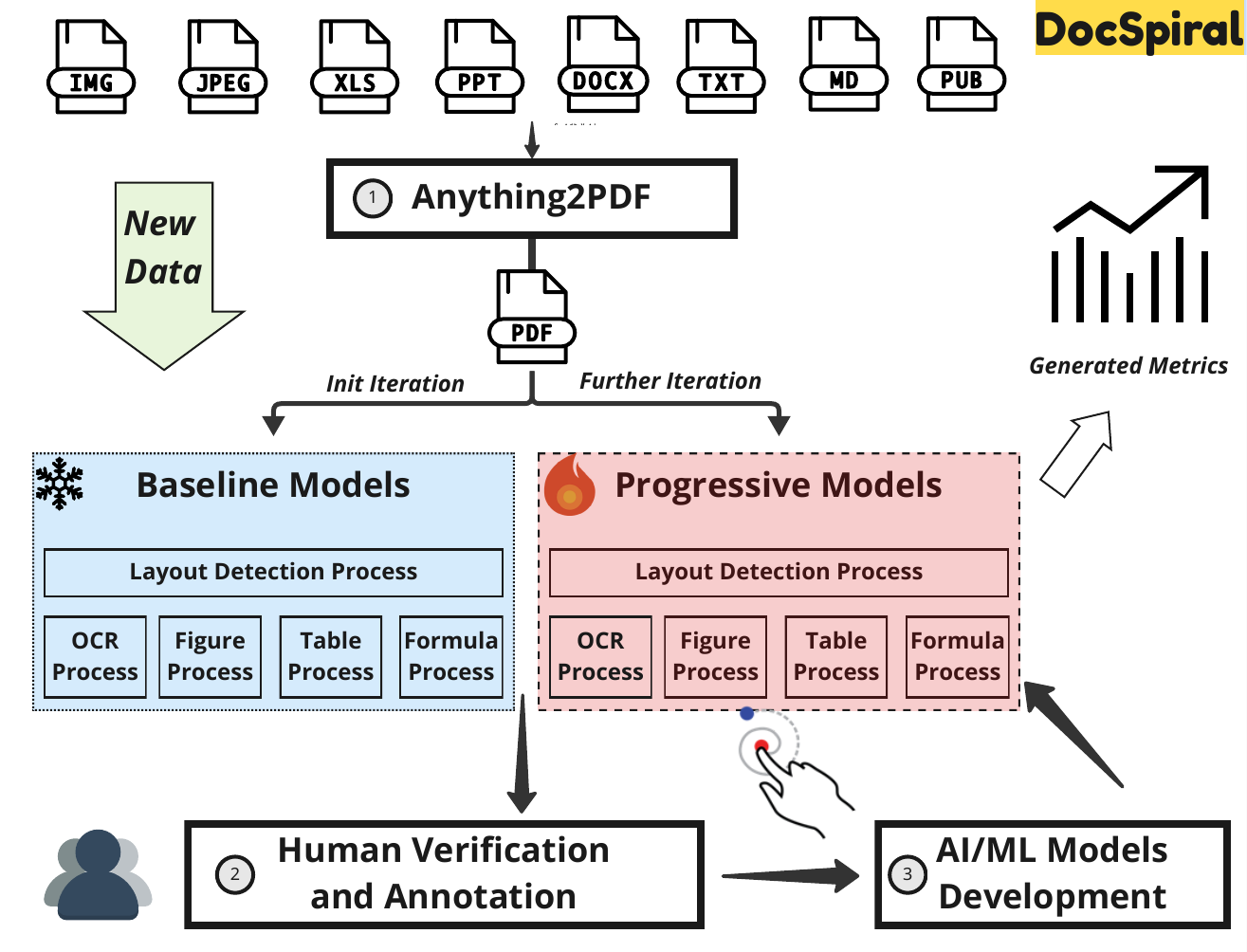}
    \caption{Our \textbf{DocSpiral} framework converts documents to PDF and processes them through iterative cycles where human verification creates annotations that improve AI/ML models, reducing effort and enhancing performance within each iteration.}
    \label{fig:framework}
    \vspace{-1em}
\end{figure}

\begin{table*}[hbt]
\centering
\scriptsize
\renewcommand{\arraystretch}{1.2}
\caption{Comparison of Document Annotation Tools. Due to the emergence of LLM and RAG technologies still being a recent development, tools supporting figure, formula, and table understanding capabilities remain scarce. (\textit{Ann.~$\Rightarrow$~Annotation, Conv.~$\Rightarrow$~Conversion: transforming data from image to another while preserving complete factual content without interpretation, Und.~$\Rightarrow$~Understanding: generating descriptive text based on a given image, involving interpretation, meaning inference, pattern recognition, and subjective judgment about data implications.)}}
\resizebox{\linewidth}{!}{%
\begin{tabular}{|l|c|c|c|c|c|c|c|c|c|c|}
\hline
\multirow{2}{*}{\textbf{Tool}} & \multirow{2}{*}{\textbf{Reference}} & \multirow{2}{*}{\textbf{Open Access}} & \multirow{2}{*}{\textbf{Layout Ann.}} & \multirow{2}{*}{\textbf{OCR Ann.}} & \multicolumn{2}{c|}{\textbf{Figure}} & \multicolumn{2}{c|}{\textbf{Formula}} & \multicolumn{2}{c|}{\textbf{Table}} \\
\cline{6-11}
& & & & & \textbf{Conv.} & \textbf{Und.} & \textbf{Conv.} & \textbf{Und.} & \textbf{Conv.} & \textbf{Und.} \\
\hline
ABBYY FineReader & \cite{abbyyfineread} & \cellcolor{red!30}No &  \cellcolor{red!30}$\times$& \cellcolor{green!30}$\checkmark$ & \cellcolor{red!30}$\times$ & \cellcolor{red!30}$\times$ & \cellcolor{red!30}$\times$ & \cellcolor{red!30}$\times$ & \cellcolor{green!30}$\checkmark$ & \cellcolor{green!30}$\checkmark$ \\
Transkribus & \cite{transkribus} & \cellcolor{red!30}No & \cellcolor{red!30}$\times$ & \cellcolor{green!30}$\checkmark$ & \cellcolor{red!30}$\times$ & \cellcolor{red!30}$\times$ & \cellcolor{red!30}$\times$ & \cellcolor{red!30}$\times$ & \cellcolor{green!30}$\checkmark$ & \cellcolor{red!30}$\times$ \\
Coco Annotator & \cite{cocoannotator} & \cellcolor{green!30}Yes & \cellcolor{green!30}$\checkmark$ & \cellcolor{red!30}$\times$ & \cellcolor{red!30}$\times$ & \cellcolor{red!30}$\times$ & \cellcolor{red!30}$\times$ & \cellcolor{red!30}$\times$ & \cellcolor{red!30}$\times$ & \cellcolor{red!30}$\times$ \\
PDFAnno & \cite{shindo-etal-2018-pdfanno} & \cellcolor{green!30}Yes & \cellcolor{red!30}$\times$  & \cellcolor{green!30}$\checkmark$ & \cellcolor{red!30}$\times$ & \cellcolor{red!30}$\times$ & \cellcolor{red!30}$\times$ & \cellcolor{red!30}$\times$ & \cellcolor{red!30}$\times$ & \cellcolor{red!30}$\times$ \\
Label Studio & \cite{LabelStudio} & \cellcolor{purple!30}Partially & \cellcolor{green!30}$\checkmark$ & \cellcolor{green!30}$\checkmark$ & \cellcolor{red!30}$\times$ & \cellcolor{red!30}$\times$ & \cellcolor{red!30}$\times$ & \cellcolor{red!30}$\times$ & \cellcolor{red!30}$\times$ & \cellcolor{red!30}$\times$ \\
PPOCRLabelv2 & \cite{ppocrlabel} & \cellcolor{green!30}Yes & \cellcolor{green!30}$\checkmark$ & \cellcolor{green!30}$\checkmark$ & \cellcolor{red!30}$\times$ & \cellcolor{red!30}$\times$ & \cellcolor{red!30}$\times$ & \cellcolor{red!30}$\times$ & \cellcolor{green!30}$\checkmark$ & \cellcolor{green!30}$\checkmark$ \\
PAWLS & \cite{neumann-etal-2021-pawls} & \cellcolor{green!30}Yes & \cellcolor{green!30}$\checkmark$ & \cellcolor{red!30}$\times$ & \cellcolor{red!30}$\times$ & \cellcolor{red!30}$\times$ & \cellcolor{red!30}$\times$ & \cellcolor{red!30}$\times$ & \cellcolor{red!30}$\times$ & \cellcolor{red!30}$\times$ \\
Tagtog & \cite{tagtog} & \cellcolor{red!30}No & \cellcolor{red!30}$\times$ & \cellcolor{green!30}$\checkmark$ & \cellcolor{red!30}$\times$ & \cellcolor{red!30}$\times$ & \cellcolor{red!30}$\times$ & \cellcolor{red!30}$\times$ & \cellcolor{red!30}$\times$ & \cellcolor{red!30}$\times$ \\
Prodigy & \cite{prodigypdf} & \cellcolor{red!30}No & \cellcolor{green!30}$\checkmark$ & \cellcolor{red!30}$\times$ & \cellcolor{red!30}$\times$ & \cellcolor{red!30}$\times$ & \cellcolor{red!30}$\times$ & \cellcolor{red!30}$\times$ & \cellcolor{red!30}$\times$ & \cellcolor{red!30}$\times$ \\
Callico & \cite{kermorvant2024callicoversatileopensourcedocument} & \cellcolor{red!30}No & \cellcolor{red!30}$\times$ & \cellcolor{green!30}$\checkmark$ & \cellcolor{green!30}$\checkmark$ & \cellcolor{green!30}$\checkmark$ & \cellcolor{red!30}$\times$ & \cellcolor{red!30}$\times$ & \cellcolor{red!30}$\times$ & \cellcolor{red!30}$\times$ \\
\textbf{DocSpiral} & Ours & \cellcolor{green!30}Yes & \cellcolor{green!30}$\checkmark$ & \cellcolor{green!30}$\checkmark$ & \cellcolor{green!30}$\checkmark$ & \cellcolor{green!30}$\checkmark$ & \cellcolor{green!30}$\checkmark$ & \cellcolor{green!30}$\checkmark$ & \cellcolor{green!30}$\checkmark$ & \cellcolor{green!30}$\checkmark$ \\
\hline
\end{tabular}%
}
\label{tab:annotation_tools_comparison}
\vspace{-1em}
\end{table*}

Most existing document processing frameworks~\cite{faysse2025colpali, shen2021layoutparser, mineru} rely on general purpose pipelines that convert raw documents into machine-readable semi-structured formats~(e.g. markdown, JSON) suitable for machine consumption. 
However, these pipelines face significant challenges when applied to domain-specific document collections, which often contain specialized terminology, unique layouts, and field-specific visual elements such as maps~\cite{zhao2024retrievalaugmentedgenerationrag,riedler2024textoptimizingragmultimodal,SurveyRAGLLM}. 
\textbf{Firstly}, traditional document processing systems often struggle to extract information accurately from such complex sources, creating barriers to knowledge utilization in fields such as geoscience, and healthcare~\cite{zhu2024realmragdrivenenhancementmultimodal}. 
The high variability and complexity of domain-specific documents necessitate human expertise to guide and refine automated processing systems. 
\textbf{Secondly}, in specialized domains, a significant portion of valuable documents exist as scanned PDFs rather than digital formats. For example, Western Australia's Mineral Exploration Reports\footnote{\url{https://www.dmp.wa.gov.au/WAMEX-Minerals-Exploration-1476.aspx}}, dating back to 1888, consist primarily of handwritten and printed documents that were later scanned into PDFs~\cite{wamex}. 
This poses challenges for automated processing, as these documents require layout analysis, optical character recognition~(OCR), and figure/table/formula processing before they can be utilised in AI-driven applications.
\textbf{Thirdly}, existing annotation tools have significant limitations for document processing tasks. Classical tools such as COCO Annotator~\cite{cocoannotator} were primarily designed for image annotation and lack optimization.
Although PAWLS~\cite{neumann-etal-2021-pawls} offers more specialized PDF labeling capabilities, it still suffers from rigid annotation schema. Currently, there is no comprehensive document annotation system that can efficiently support the entire document annotation pipeline.
Due to the diversity of output structures for figure/table/formula processing, we also need such systems capable of addressing the dynamic complexity of these tasks through features like dynamic annotation form generation.

To address these challenges, we introduce \textbf{DocSpiral}, the first Human-in-the-Spiral assistive document annotation platform designed to facilitate domain-specfic document processing.
As shown in Figure~\ref{fig:framework}, our system first converts various document formats to standardized PDF format through the \textbf{Anything2PDF} module.
This unified format enables integration with layout analysis models like \textit{DocLayout-YOLO} as \textbf{Baseline Models}, which predict bounding boxes using a generic layout schema~\cite{zhao2024doclayoutyoloenhancingdocumentlayout}.
Based on bounding-box labels, specialized downstream processing~(OCR, Figure/Table/Formula processing) is triggered using the corresponding \textbf{Baseline Models}.
Our human-in-the-spiral approach, supported by an interactive interface, allows experts to review, verify, and annotate model output. 
The annotated data are then used to train or fine-tune to obtain \textbf{Progressive Models} that better meet user requirements.
Unlike existing tools such as COCO Annotator~\cite{cocoannotator}, our system leverages pre-trained models for initial annotation, significantly reducing the manual labeling workload. 
Users can focus primarily on corrections through an intuitive web-based interface, which leads to at least 41\% time reduction, and in some cases up to 75\%.

Our work makes three key contributions:
\vspace{-0.5em}
\begin{itemize}[noitemsep, left=0pt]
    \item \textbf{Comprehensive Document Annotation System} – We develop the first~(as shown in Table~\ref{tab:annotation_tools_comparison}) full-featured annotation system that supports the entire document processing pipeline, from layout detection and OCR to tables, figures, and formulas conversion and understanding tasks. Our flexible and customizable annotation schema design accommodates the complexity and diversity of layout, figure/formula/table processing tasks.
    \item \textbf{Assisted Spiral Improvement Framework} – We introduce an iterative, human-in-the-spiral approach where human verification and model training reinforce each other over successive cycles. This process progressively reduces annotation effort while improving model performance.
    \item \textbf{Open and Deployable Solution} – We make DocSpiral freely accessible to researchers while also offering deployable solutions for organizations with privacy constraints, thereby removing barriers to LLM adoption in specialized domains.
\end{itemize}
\vspace{-0.5em}

\section{Related Work}

\paragraph{Existing annotation tools}

Comprehensive document annotation requires support for various tasks, including \textbf{\textit{Layout detection}} outputs bounding boxes with category labels~(content, title, figure, table, formula, footnote, etc); \textbf{\textit{OCR}} requires accurate text transcriptions of segmented images; \textbf{\textit{Table processing}} includes both structural conversion (to LaTeX~\cite{xia2024docgenome}, HTML~\cite{Wan_2024_CVPR}, JSON~\cite{table2json}) and semantic understanding~\cite{zhao2024tabpedia} for RAG systems as explained in Table~\ref{tab:annotation_tools_comparison}; \textbf{\textit{Formula processing}} is similar to table processing with structural conversion typically outputting LaTeX~\cite{xia2025latexnet}; \textbf{\textit{Figure processing}} prioritizes understanding visual elements over conversion due to representation diversity and difficulty~\cite{NEURIPS2024_0d97fe65}. 
Tools such as LayoutParser~\cite{shen2021layoutparser}, MinerU~\cite{mineru}, and Docling~\cite{Docling} provide the ability to integrate with parts or all of these specialized models to build an end-to-end pipeline; however, when errors occur, these tools lack mechanisms that allow users to fix specific problems or improve individual models at intermediate stages.

No existing annotation tool fully addresses these needs, as illustrated in Table~\ref{tab:annotation_tools_comparison}, particularly for semantic understanding annotation of formulas, tables, and figures. Commercial solutions such as Tagtog~\cite{tagtog}, Callico~\cite{kermorvant2024callicoversatileopensourcedocument}, and Prodigy~\cite{prodigypdf} offer partial capabilities. Although there are specialized tools for individual tasks~\cite{huynh2022heuristics, gipplabAnnoMathTeX}, the research community lacks a unified system that integrates these capabilities into a cohesive pipeline that facilitates human intervention for error correction and iterative model improvement throughout the entire document processing workflow, to produce structured high-quality outputs from unreconstructed data formats.

\paragraph{Human role in annotation systems}
Traditional annotation pipelines follow a \emph{human-off-the-loop} paradigm, where annotators exhaustively label data offline before model training or fine-tuning~\cite{LabelStudio,kermorvant2024callicoversatileopensourcedocument,prodigypdf,neumann-etal-2021-pawls}. While effective, this approach is labour-intensive and impractical for large or continuously evolving datasets~\cite{wu2022survey,pena2024continuous}.

Instead, our document processing framework shifts towards a \emph{human-in-the-loop} approach~\cite{nahavandi2017trusted}, using baseline models or LLMs for assistive annotations—such as figure understanding or layout detection—while humans intervene selectively for validation and correction.

Building on this, we introduce the \emph{human-in-the-spiral} framework, where new data is first processed by prior models, then undergoes targeted expert review, followed by iterative model enhancement (as shown in Figure~\ref{fig:framework}). This positive feedback loop improves model performance in an upward spiral while minimizing manual annotation.
\section{System design and implementation}
\subsection{Requirement analysis}
We present \textbf{DocSpiral}, a web-based document processing pipeline initiated from document layout detection~\cite{huang2022layoutlmv3pretrainingdocumentai}. 
Its primary objective is to enable efficient review and annotation of model outputs, generating high-quality annotated data for iterative models improvement. This approach enhances downstream tasks, such as knowledge graph construction and RAG applications, in specialized domains like geology and healthcare, where valuable documents are often paper-based, or in scanned images with a rich mix of multimodal contents, such as photos, maps, charts, and tables.

To ensure usability and adaptability, the system minimizes human effort, incorporates automatic evaluation metrics dashboard generation, and maintains a modular structure for extensibility.
It supports agile methodologies, allowing researchers to rapidly develop and refine different document processing models in response to the fast paced GenAI challenges.
As shown in Figure~\ref{fig:framework}, \textbf{DocSpiral} consists of three integrated modules:

\vspace{-0.5em}
\begin{itemize}[noitemsep, left=0pt]
    \item \textbf{Anything2PDF} converts diverse formats (Word, PowerPoint, Excel, images, text, ebooks, Markdown) into standardized PDFs, creating a unified processing foundation.
    \item \textbf{Annotation Interface} provides a web-based platform to annotate layout detection, OCR, tables, figures, and formulas outputs.
    \item \textbf{AI/ML Models Enhancement} supports continuous models enhancement through data download and result submission API endpoints, and automatic evaluation metrics dashboard generation~(latency and accuracy).
\end{itemize}

\begin{figure}[hbt]
    \centering
    \includegraphics[width=\linewidth]{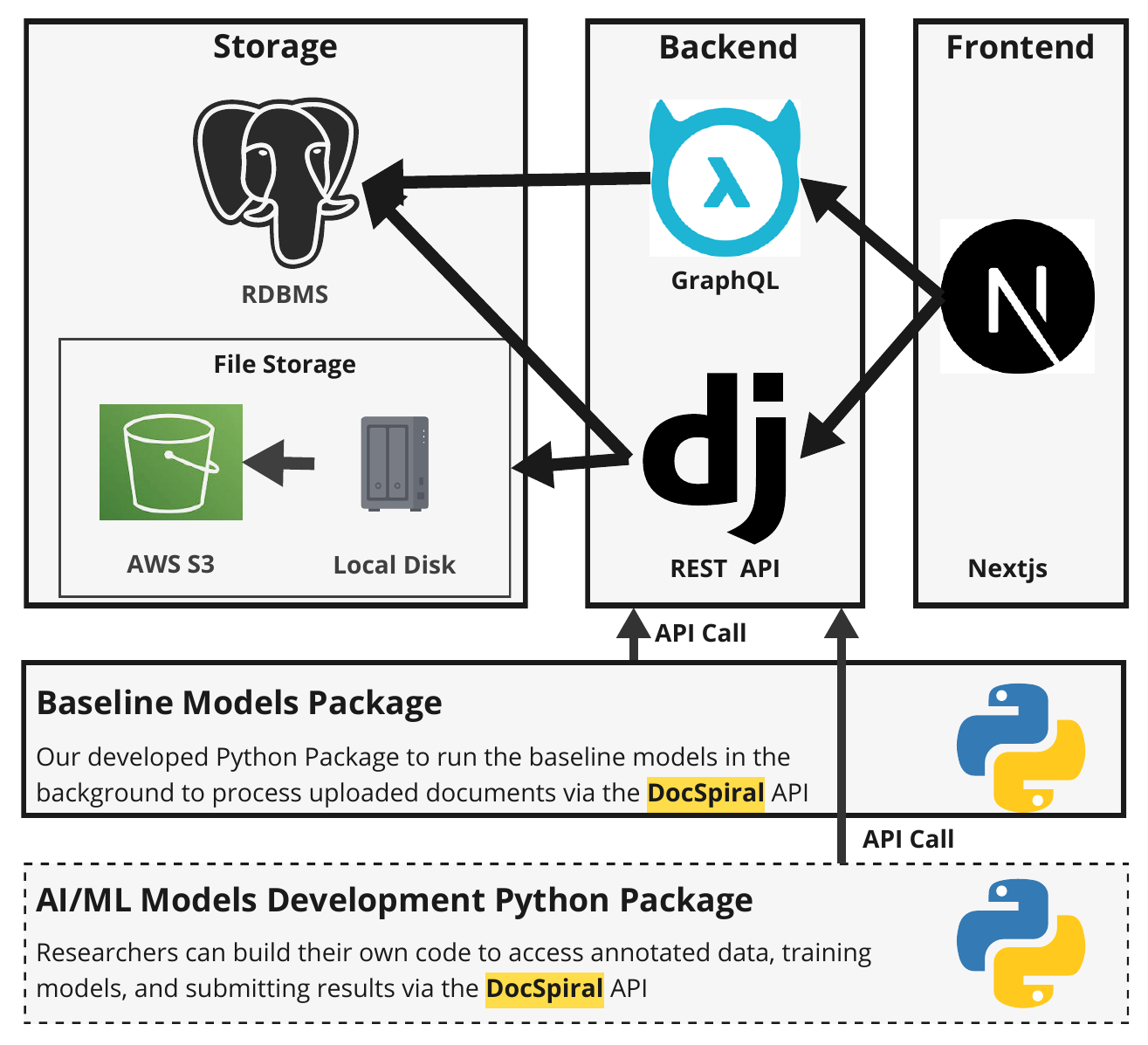}
    \caption{System Architecture Overview for \textbf{DocSpiral}}
    \label{fig:docspiraldesign}
    \vspace{-1em}
\end{figure}


\subsection{System design}
The software stack of \textbf{DocSpiral} is illustrated in Figure~\ref{fig:docspiraldesign}, which consists of three main layers:
\vspace{-0.5em}
\begin{itemize}[noitemsep, left=0pt]
  \item \textbf{Frontend:} Built with React/Next.js\footnote{\url{https://nextjs.org/}}, providing a responsive web-based user interface.
  \item \textbf{Backend:} Consists of two frameworks: (1) Hasura\footnote{\url{https://hasura.io/}} to generate GraphQL endpoints, enabling rapid feature development and supporting \textbf{real-time} collaboration among annotators; (2) Django\footnote{\url{https://www.djangoproject.com/}} for database migration management, user authentication, RESTful endpoints, and AWS S3\footnote{\url{https://aws.amazon.com/}} integration, etc.
  \item \textbf{Storage:} Uses PostgreSQL\footnote{\url{https://www.postgresql.org/}} for metadata, user management, annotations, evaluation metrics, and task tracking, while document files are securely stored in a private S3 bucket.
\end{itemize}
\vspace{-0.5em}
Additionally, we have developed a Python package that interacts with the platform via API calls to run baseline models and report results. Researchers can extend these functionalities by creating their own packages to download documents and annotated data, train models, perform inference using the \textbf{Progressive Models}, and submit results through the RESTful endpoints. The system is deployed in the Amazon Web Services~(AWS) for scalable and secure infrastructure.

\subsection{Implementation}
\begin{figure}[hbt]
    \centering
    \includegraphics[width=\linewidth]{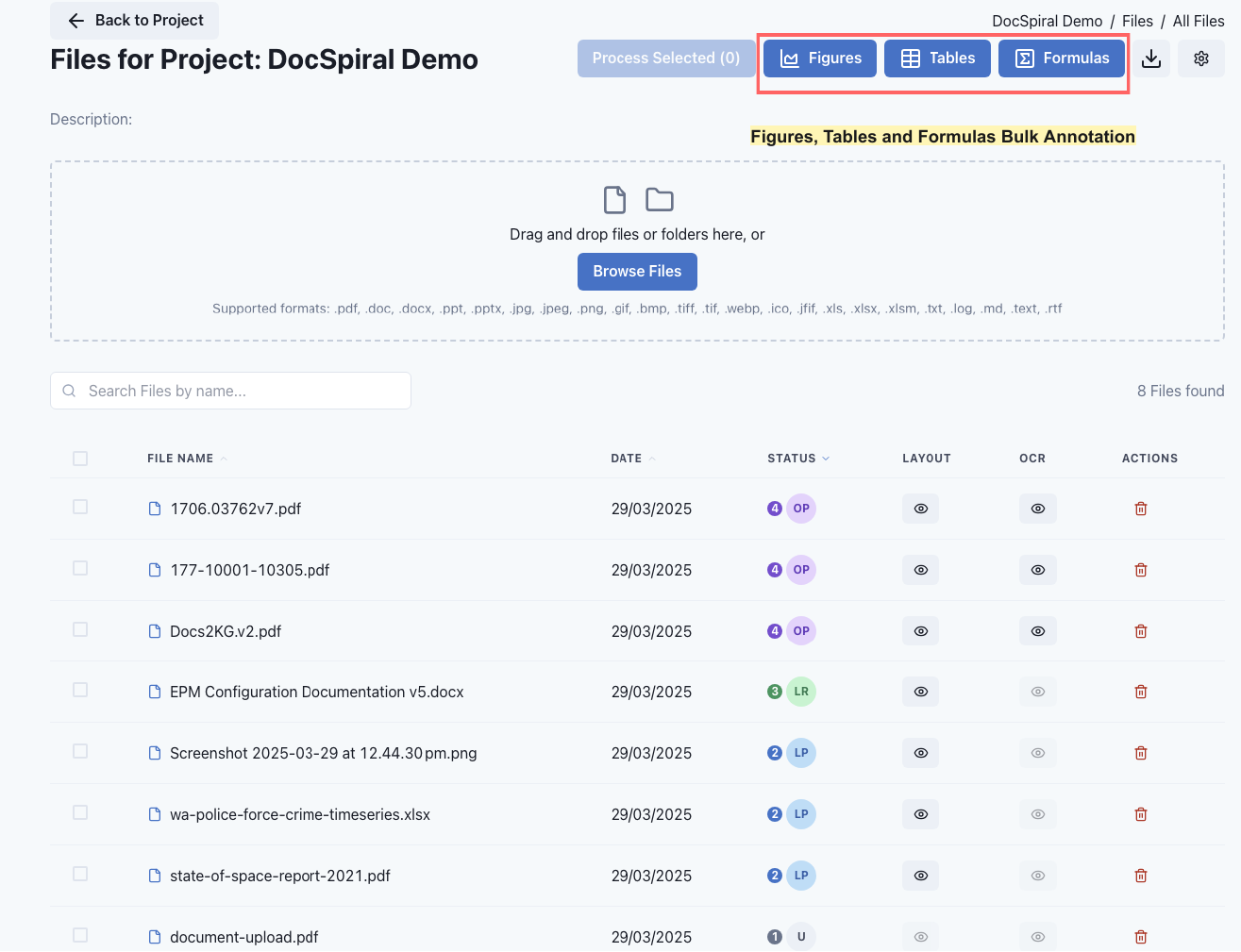}
    \caption{Documents upload and management interface, users can drag and drop allowed format documents or zipped files. Files will be uploaded to S3 bucket, and downstream tasks will be triggered.}
    \label{fig:documentupload}
    \vspace{-0.5em}
\end{figure}
\noindent\textbf{a.)~Documents Uploading:} Users start by registering an account, creating a project within \textbf{DocSpiral}, and uploading documents through the interface shown in Figure~\ref{fig:documentupload}. PDF files undergo immediate layout detection process, while other formats are first converted to PDF via \textbf{Anything2PDF}. Uploaded documents can be managed within the platform, and users can track the processing progress of each document. 
The system assigns the following status values: \circled{1} \textit{Uploaded}
\circled{2} \textit{Layout detection completed}
\circled{3} \textit{Human-reviewed layout}
\circled{4} \textit{OCR and processing of figures, tables, and formulas completed}
\circled{5} \textit{Human-reviewed model outputs from previous step}.

Once a document reaches status \circled{2} or higher, the layout \textit{“eye”} icon becomes clickable for review. Similarly, the OCR column becomes interactive once the document reaches status \circled{4} or above.

\begin{figure}[hbt]
        \centering
        \includegraphics[width=\linewidth]{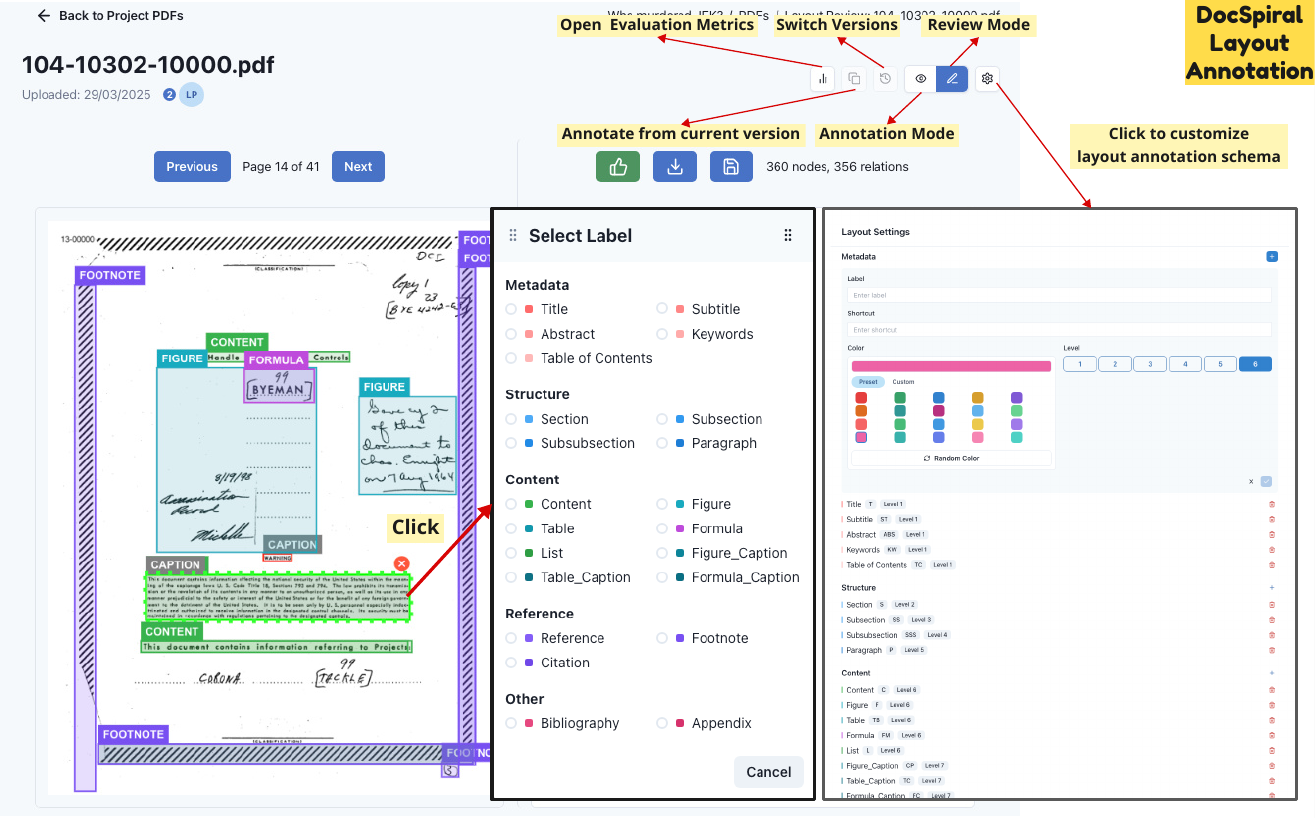}
        \caption{Layout annotation interface: user can click, add or remove bounding boxes from PDF Viewer, and assign layout labels~(middle), or customize a domain specific hierarchical layout schema~(right).}
        \label{fig:layout-annotation}
        \vspace{-1em}
\end{figure}

\noindent\textbf{b.)~Layout Detection Annotation:}
We use DocLayout-YOLO~\cite{zhao2024doclayoutyoloenhancingdocumentlayout} as baseline model for layout detection. You can speficy your own baseline model when you use \textbf{DocSpiral}.
This model treats each PDF page as an image and outputs bounding boxes for detected layout elements along with their corresponding labels. The data of the bounding box are represented as $[x_{min}, y_{min}, width, height]$ in normalized coordinates.
Document layouts vary significantly across domains, making it difficult to develop a universal layout detection model. For example, our baseline model supports only a limited set of labels—content, title, figure, table, formula, footnote—and struggles with complex layouts, such as PDF forms in hospital, you may want to define a label for patient name. To address this, our system enables domain-specific customization: (1) users can define their custom layout schema with hierarchies and (2) refine annotations by reclassifying, removing or adding bounding boxes with labels~(Figure~\ref{fig:layout-annotation}).
\begin{figure}[hbt]
    \centering
    \includegraphics[width=\linewidth]{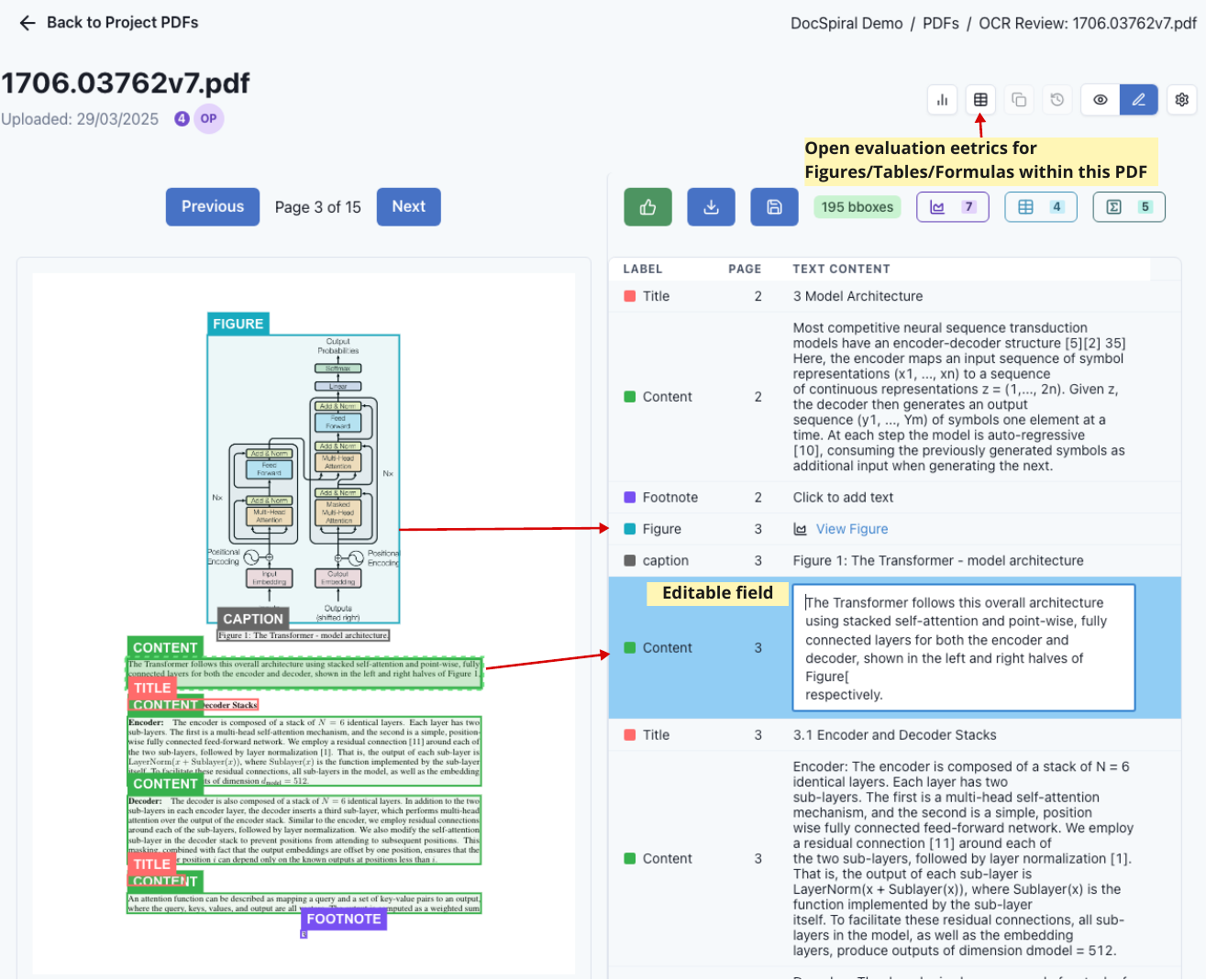}
    \caption{OCR verification and annotation interface}
    \label{fig:ocr-annotation}
\end{figure}

\noindent\textbf{c.)~OCR Annotation:}
After reviewing the layout detection results, users can save and trigger downstream processing, including the OCR process.
We use PaddleOCR~\cite{ppocrlabel} as the baseline OCR model for its strong performance, multilingual support, and ease of use. 
OCR results for each layout block appear in a table alongside their labels. The interface enables interactive navigation: clicking a bounding box in the PDF viewer scrolls the table to the corresponding row, while selecting a table row highlights the relevant section in the PDF viewer. Users can edit incorrect text directly, with changes auto-saved, as shown in Figure~\ref{fig:ocr-annotation}.

\noindent\textbf{d.)~Table, Formula and Figure Annotation:}
There are several models that process table images to output diverse formats for different purposes. We support multiple formats for table conversion~(HTML~\cite{PIX2TEXTTABLE}, LaTeX~\cite{xia2024docgenome}, JSON~\cite{table2json}) using various baselines: Pix2Text for HTML, StructEqTable for LaTeX, and a vision LLM agent for JSON extraction. For formula conversion, we output LaTeX using Pix2Text, while figure understanding leverages a vision LLM to generate descriptive text.

\begin{figure}[hbt]
   \centering
   \includegraphics[width=\linewidth]{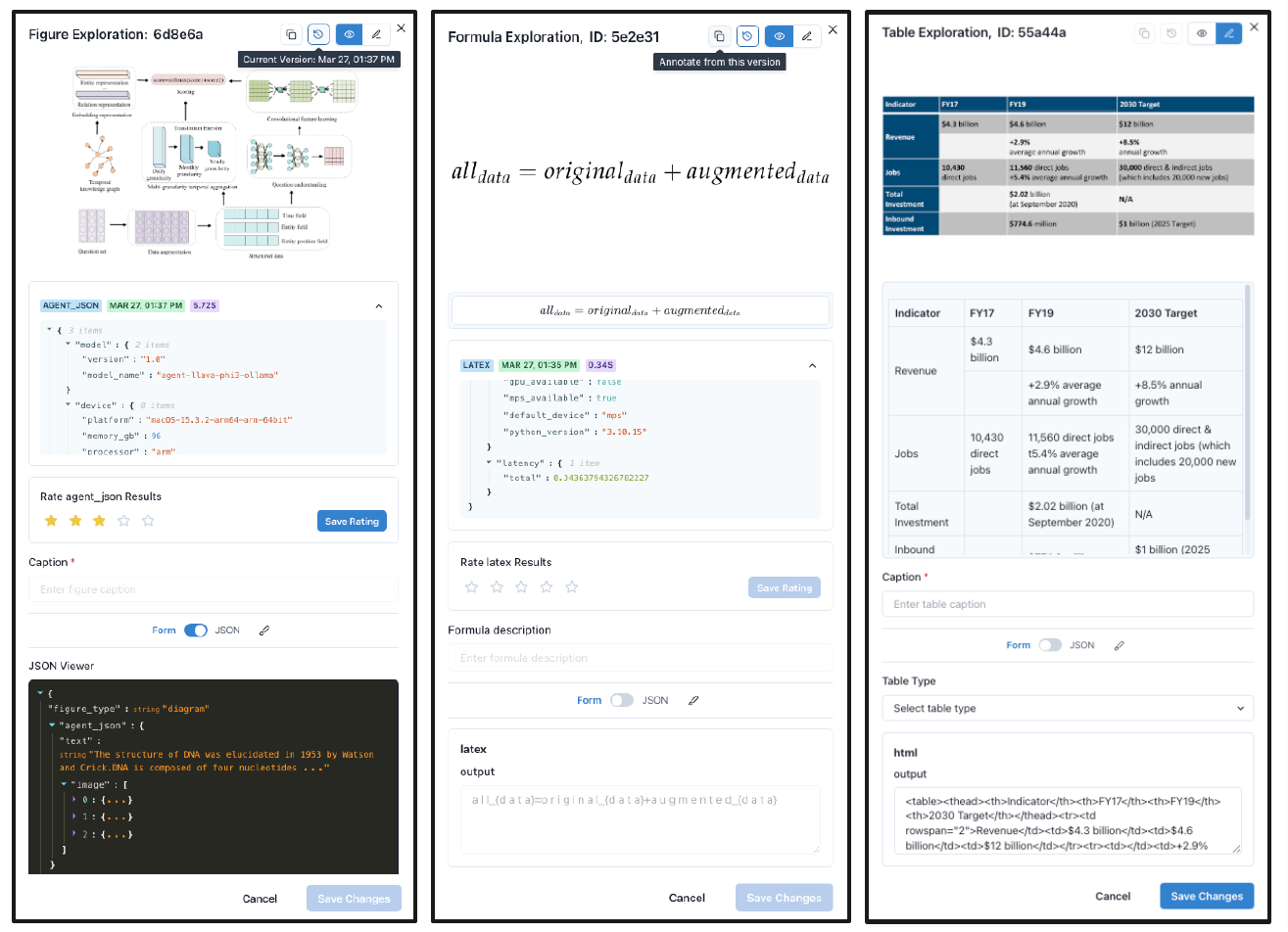}
   \caption{Figure annotation interface in \texttt{review mode} with JSON viewer (left); Formula in \texttt{review mode} showing \texttt{latex} output in form (middle); Table in \texttt{annotation mode} with editable \texttt{output} field from \texttt{html} model using schema-generated form (right).}
   \label{fig:table-annotation}
\vspace{-1em}
\end{figure}

\begin{figure}[hbt]
       \centering
       \includegraphics[width=\linewidth]{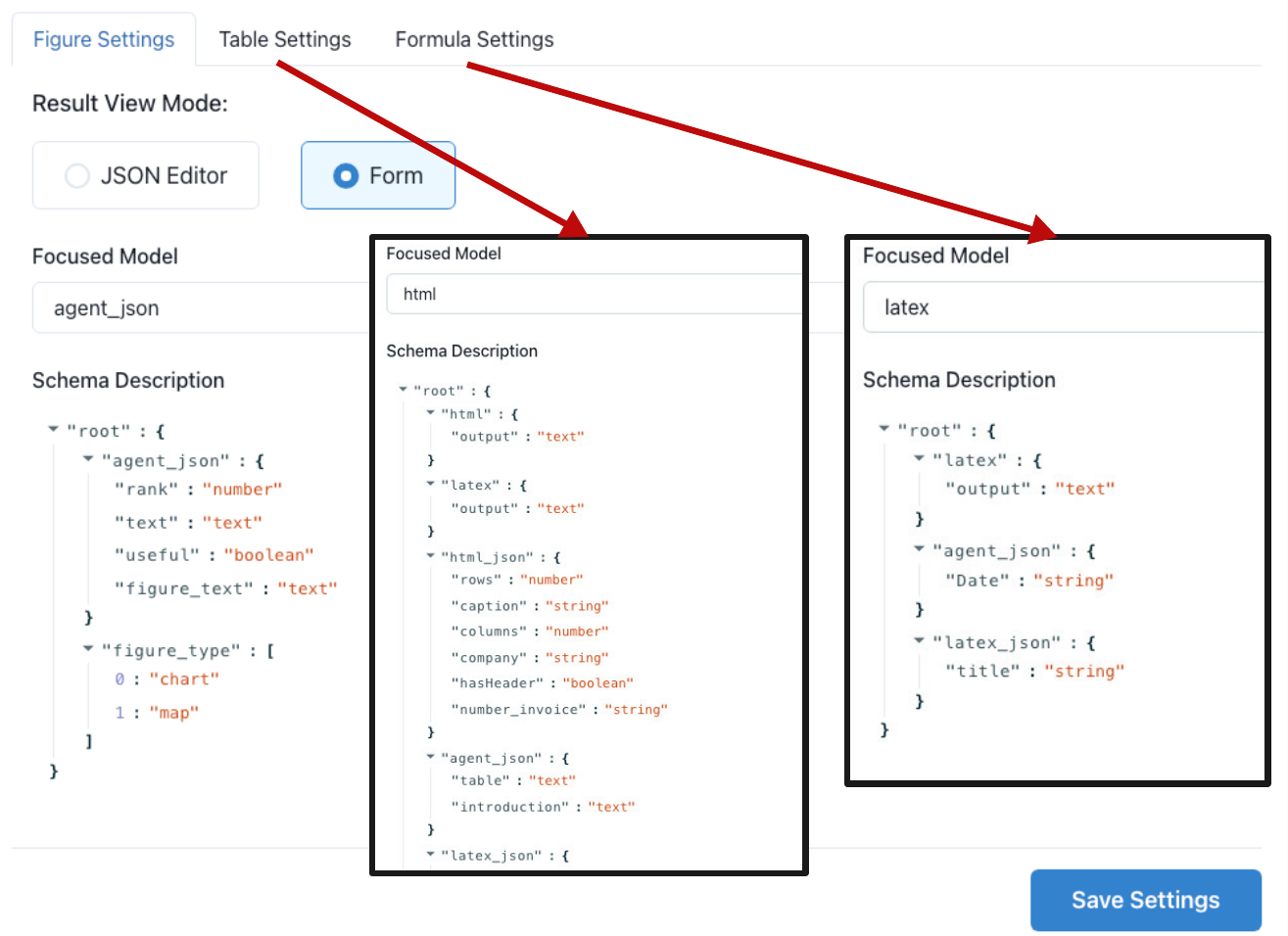}
       \caption{The settings interface allows configuring form schema and selecting a \texttt{Focused Model}. Outputs display as raw \texttt{JSON}~(Figure~\ref{fig:table-annotation} left) or as structured \texttt{Form} (Figure~\ref{fig:table-annotation} middle/right) with fields determined by the selected model's form schema.}
       \label{fig:figuretableformula_annotation}
      \vspace{-1em}
\end{figure}

When users click on a \textit{Figure}, \textit{Formula} or \textit{Table} row, a corresponding annotation interface appears~(Figure~\ref{fig:table-annotation}). Since models for different purposes require different output formats for figures, formulas, and tables, we implemented a dynamic, flexible annotation interface through \textbf{our annotation form generation feature}: users define their \texttt{Focused Model} and form schemas in settings~(Figure~\ref{fig:figuretableformula_annotation}), and the interface generates appropriate input fields based on the schema of the selected model. For example, selecting the \texttt{html} model for table annotation creates a TextArea field named \texttt{output}, while switching to \texttt{html\_json} generates input fields for rows, caption, etc. Model outputs are prepopulated when possible to reduce manual effort. The \texttt{JSON Editor} mode allows users to examine all model outputs for better observability and to inform better annotation form schema design.

For improved efficiency, \textbf{DocSpiral} supports bulk annotation of figures, tables, and formulas across projects via buttons in Figure~\ref{fig:documentupload}. Users can also upload standalone images for direct annotation without starting from PDF layout detection.

\begin{figure}[hbt]
   \centering
   \includegraphics[width=\linewidth]{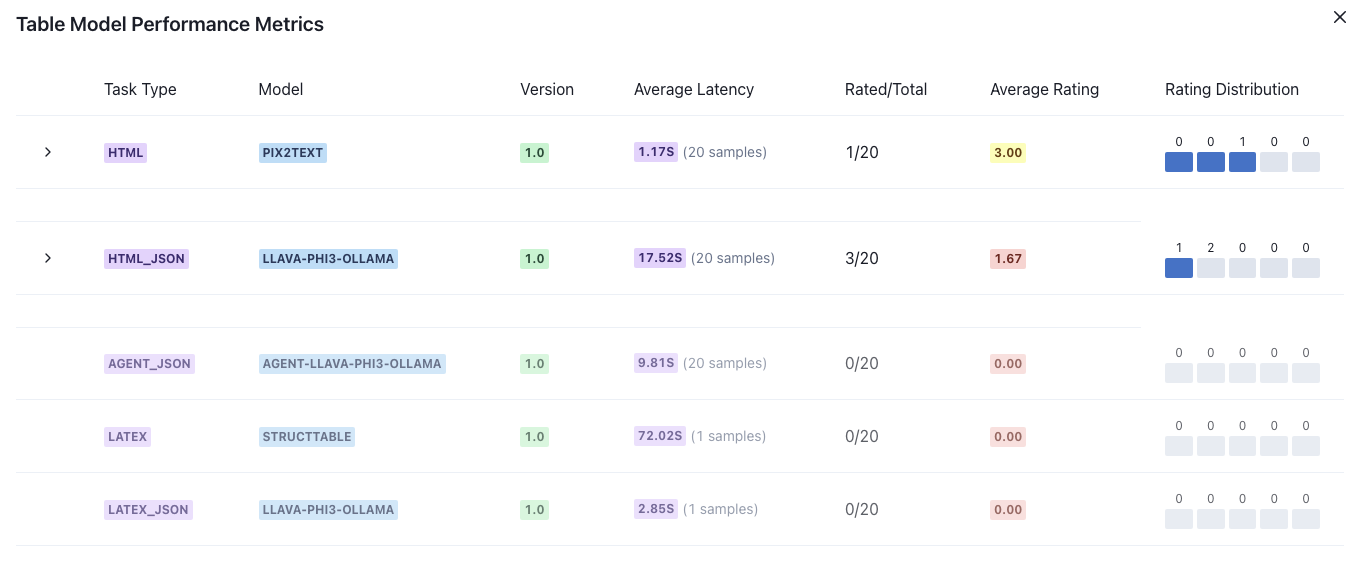}
    \caption{Table model performance dashboard displaying metrics across different output type, models, and versions. Metrics include latency, human satisfaction ratings, annotation and review progress tracking.}
   \label{fig:table-metrics}
    \vspace{-0.8em}
\end{figure}

\noindent\textbf{e.)~Metrics Dashboard Generation:}
\textbf{DocSpiral} tracks objective measures (mAP for layout detection, CER/WER for OCR) and records latency for all model runs. For subjective outputs (figures, formulas, and tables processing), we implement \textbf{human satisfaction ratings} feature to quantify model-human alignment~(Figure~\ref{fig:figuretableformula_annotation} left and middle). A centralized dashboard~(exemplified in Figure~\ref{fig:table-metrics}) aggregates these metrics to monitor model performance and annotation progress. Other evaluation metrics can be added based on user feedback.

\noindent\textbf{f.)~Model Development Support:}
Raw data (PDFs, figures, tables, and formulas) and annotations are securely accessible via authenticated RESTful endpoints, together with submission of models outputs. Detailed instructions are available from \textbf{DocSpiral} documentation.

\section{System evaluation}

We quantitatively evaluated \textbf{DocSpiral}'s efficiency through an annotation experiment with 90 diverse document pages. Baseline model-assisted annotation reduced processing time from 28.4s to 16.7s per page compared to manual annotation, yielding a 41\% overall time reduction. For low-quality scanned PDFs, time reduced by 75\%.

\begin{table}[hbt]
\centering

\caption{Faster-RCNN Training Performance}
\label{tab:model_evolution}
\scriptsize
\resizebox{\linewidth}{!}{%
\begin{tabular}{|l|c|c|c|c|}
\hline
\textbf{Metric} & \textbf{Initial} & \textbf{1st} & \textbf{2nd} & \textbf{3rd} \\
\hline
mAP (\%) & 0.053 & 0.12 & 0.21 & 0.33 \\
\hline
\end{tabular}%
}
\end{table}

We identify three promising pathways for model spiral evolution: (1) \textbf{Traditional rule-based solutions} benefit from improved observability, enabling targeted fixes such as removing footnotes in specific locations; (2) \textbf{Deep learning models} like Faster-RCNN~\cite{ren2016fasterrcnnrealtimeobject} can be fine-tuned or redesigned and trained using annotated data; (3) \textbf{Large language models (LLMs)} can be fine-tuned for better domain-specific alignment in figure, table and formula understanding. We experiment with Faster-RCNN training for layout detection over three iterative cycles, each adding 100 new pages of data, demonstrated progressive performance gains (Table~\ref{tab:model_evolution}), validating our methodology.

\section{Conclusion}
This paper presents \textbf{DocSpiral}, the first integrated assistive document annotation platform through Human-in-the-Spiral for extracting structured information from domain-specific, image-based unstructured documents.
Our system introduces three key innovations: comprehensive annotation interfaces supporting the entire document processing pipeline, a customizable hierarchical layout schema, and dynamic annotation form generation for figures, formulas, and tables. Experiments demonstrate that our approach reduces annotation time by at least 41\% while showing consistent performance gains across iterations. 
In summary, the human-in-the-spiral approach establishes an iterative cycle where human annotations and model performance reinforce each other, progressively reducing manual effort and enhancing model performance.
By making this platform freely accessible, we aim to lower barriers to AI/ML model development in document-intensive fields. We continue to improve the system based on user feedback and plan to open-source it once functionality stabilizes.

\newpage
\bibliography{ref}

\end{document}